\documentclass[twoside,leqno,11pt]{article}  

\usepackage{amsmath,amssymb}
\usepackage{amsthm}
\usepackage[dvips]{hyperref}
\usepackage{bm}
\usepackage[english]{babel}

\newcommand{\be}{\begin{equation}}
\newcommand{\ee}{\end{equation}}
\newcommand{\ef}[1]{\,#1}

\newcommand{\lrceil}[1]{\lceil #1 \rceil}

\newcommand\rd{{\mathrm d}}

\newcommand\peq{\subseteq}

\newcommand{\reof}{\mathfrak{Re}}

\newcommand{\ba}{{\bm \alpha}}
\newcommand{\z}{\zeta}
\newcommand{\e}{\epsilon}

\renewcommand{\a}{\alpha}

\newcommand{\cX}{\mathcal{X}}
\newcommand{\cN}{\mathcal{N}}
\newcommand{\bn}{{\bm n}}
\newcommand{\bv}{{\bm v}}
\newcommand{\bbN}{\mathbb{N}}
\newcommand{\bbR}{\mathbb{R}}
\newcommand{\bbC}{\mathbb{C}}
\newcommand{\bbZ}{\mathbb{Z}}

\newcommand{\myitem}[1]{\rule{0pt}{13pt}\quad #1)\ }

\newcommand{\ptil}{\;\;\makebox[0pt][c]{\raisebox{-3pt}{$\sim$}}\makebox[0pt][c]{\raisebox{1pt}{$+$}}\;\;}

\newcommand{\stisk}[2]{\genfrac{\{}{\}}{0pt}{}{#1}{#2}}

\newcommand{\fracdsline}[2]{\left. \left( \displaystyle{#1} \right)
  \middle/ \left( \displaystyle{#2} \right) \right.}

\newcommand{\cO}{\mathcal{O}}
\newcommand{\tT}{\Theta}
\newcommand{\cF}{\mathcal{F}}

\newtheorem{theorem}{Theorem}[section]

\newtheorem{proposition}[theorem]{Proposition}

\begin{document}

\title{\Large Linear-time generation of specifiable combinatorial
  structures: \\
general theory and first examples}

\author{
{\large Fr\'ed\'erique Bassino \qquad Andrea Sportiello}\\
\rule{0pt}{16pt}%
\normalsize{LIPN, CNRS UMR 7030, Universit\'e Paris Nord,}\\
\normalsize{99 av.\ J.-B.\ Cl\'ement, 93430 Villetaneuse, France.}\\
\normalsize{{\it e-mail:} \quad {\tt bassino},\quad
{\tt sportiel{}lo@l{}ipn.univ-paris13.fr}}
}

\date{July 5, 2013}


\maketitle

\begin{abstract}
\noindent
Various specifiable combinatorial structures, with $d$ extensive
parameters, can be exactly sampled both by the recursive method, with
linear arithmetic complexity if a heavy preprocessing is performed,
or by the Boltzmann method, with average complexity $\Theta(n^{1+d/2})$.

We discuss a modified recursive method, crucially based on the
asymptotic expansion of the associated saddle-point integrals, which
can be adopted for a large number of such structures
(e.g.\ partitions, permutations, lattice walks, trees, random graphs,
all with a variety of prescribed statistics and/or constraints). The
new algorithm requires no preprocessing, still it has linear
complexity on average. In terms of bit complexity, instead of the
arithmetic one, we only have extra logarithmic factors.  For many
families of structures, this provides, at our knowledge, the only
known quasi-linear generators.

We present the general theory, and detail a specific example: the
partitions of $n$ elements into $k$ non-empty blocks, counted by the
Stirling numbers of the second kind. These objects are involved in the
exact sampling of minimal automata with prescribed alphabet size and
number of states, which is thus performed here with average
$\Theta\big( n \ln n \big)$ bit complexity, outbreaking all previously
known $\Theta\big( n^{3/2} \big)$ algorithms.
\end{abstract}

\noindent {\it Keywords:}
Random combinatorial structures,
Random generation,
Recursive method,
Random minimal automata.

\thispagestyle{empty}

\newpage

\setcounter{page}{1}

\section{Introduction}
\label{sec.intro}

This paper deals with the exact sampling of random combinatorial
structures $X$, from measures on statistical ensembles with multiple
size parameters, $X \in \cX_{\bn}$, $\bn = (n_1, \ldots, n_d) \in
\bbN^d$.  We address the case in which the structures have a
\emph{combinatorial specification} 
%
(see \cite{flaj}, sec.~I.2 and references therein), i.e.\ are
described in terms of elementary
constructors
(disjoint union, 
cartesian product, 
sequence, 
set, multiset,
cycle, 
\ldots),
a situation in which, under some mild further hypotheses, there exist
already two general algorithmic strategies: the \emph{recursive}
\cite{flajRecuZVC, wilf} and the \emph{Boltzmann} methods
\cite{flajBS, FFPiv}. Setting $N=\sum_j n_j$ the sum of the size
parameters,
the recursive method has bit complexity $\tT(N)$ or $\tT(N \log N)$ in
most cases,\footnote{We say \emph{quasi-linear} to denote the two
  possibilities altogether.} whenever the coefficients of the
generating function have explicit fast-computable formulas. However,
if this is not the case, it has only a poor
$\tT(N^{d+1})$\;\footnote{Here and in the whole paper, we neglect $\ln
  N$ factors in complexity, when this is $\tT(N^{\gamma})$,
  $\gamma>1$.}
time and space complexity.\;\footnote{To some extent, one can reduce
  the space complexity, while degrading the time complexity, see later on.}
On the other side, the Boltzmann method has a time complexity
$\tT(N^{d/2+1})$ on average, quasi-linear space complexity, and a
wider range of applicability. A natural goal is to fill this gap, and
provide a `mixed' algorithm that achieves a quasi-linear space and
(average-)time complexity, with no preprocessing, essentially in every
context for which a Boltzmann sampling is available. Within this paper
we shall require an extra property, pertinent to the recursive method,
namely that we have a linear recursion at the level of the generating
functions that implies an algorithmic step-by-step construction of the
structure (in particular, to a certain extent, the recursion must have
non-negative coefficients).  This is in fact quite often the case for
objects within the symbolic method framework.

For various special cases of combinatorial structures,
linear or quasi-linear algorithms have been designed. We mention in
particular, as prototype examples, the Remy algorithm \cite{remy}, for
generating random planar binary trees of a given size, and a recent
extension
\cite{bachbodjacq} for unary-binary trees.
These algorithms are elegant, and intrinsically combinatorial. 
The drawback is that they are rare gems, and exist only for very few
specific problems.  On the contrary, the strategy we present here
aims to be quite general, and extend to weighted objects with a
minimal amount of extra work.

Within the theory developed here, and supplied with the (easy)
verification of the conditions in Section \ref{sec.summconds}, one can
produce quasi-linear algorithms for sampling: 
(1)~partitions of a set, constrained to the number of blocks, and
possibly the set of allowed cardinalities (that we discuss here in
detail);\\
(2)~per\-mu\-ta\-tions, constrained to the number of cycles, and possibly
the set of allowed cycle lengths;
(3)~walks and directed walks, constrained to their endpoints, and to
other statistics, e.g., in $\bbZ^2$, the area encircled by the path
(these further statistics make the problem non-trivial); (4) Various
families of trees, e.g.\ with prescribed number of nodes for each
degree\ldots \
%
In particular, in conjunction with the results in 
\cite{BasNic, usMinAuto}, our algorithm for the first example implies
the quasi-linear uniform generation of random $n$-state minimal
automata over a $k$-symbol alphabet, for any $k \geq 2$.

\section{Two examples}

Before setting up a general theory, let us illustrate with
some specific examples how the
`classical' recursive method works,
and why one should expect that our enhancement
is feasible.  Our first example is `too easy' for us to improve on
previous complexity: sampling a random directed walk on $\bbN^2$, from
$(0,0)$ to $(n,m)$. There are $\binom{n+m}{n}$ such walks, satisfying
the binomial relation
\be
\label{eq.binoms}
\binom{n+m}{n} = \binom{n+m-1}{n-1} 
+ \binom{n+m-1}{n}
\ef.
\ee
We stress the fact, important at our aims, that this relation can be
rephrased into an algorithmic construction, based on a branching
procedure: if one could sample uniformly from the ensembles
$\cX_{n-1,m}$ and $\cX_{n,m-1}$, and could efficiently toss a biased
coin with parameter $p_{n,m}= \binom{n+m-1}{n-1} / \binom{n+m}{n}$,
then one could sample uniformly from $\cX_{n,m}$, by first tossing the
coin, then, depending from the result, appending ``north'' or ``east''
in the list of steps, and sampling uniformly from the first or the
second ensemble, respectively.
Let $T^{\rm step}_{E,N}$ the complexity needed to add one (east or
north) step to our constructed object. This is thus a constant,
independent of $n$ and $m$, and the overall average and worst-case
complexities satisfy the associated linear relations
\begin{subequations}
\label{eqs.complerec}
\begin{align}
\begin{split}
T^{\rm aver.}_{n,m} 
&=
T^{\rm coin}_{n,m} 
+
\left(
p_{n,m}
(
T^{\rm aver.}_{n-1,m} 
+ T^{\rm step}_E
)
+
(1-p_{n,m})
(
T^{\rm aver.}_{n,m-1} 
+ T^{\rm step}_N
)
\right)
\ef;
\end{split}
\\
T^{\rm worst}_{n,m} 
&=
T^{\rm coin}_{n,m} 
+
\max
\left(
T^{\rm worst}_{n-1,m}
+ T^{\rm step}_E
,
T^{\rm worst}_{n,m-1} 
+ T^{\rm step}_N
\right)
\ef.
\end{align}
\end{subequations}
We can thus recursively push our calculation of complexity to the
sole delicate point, the complexity of producing the properly-biased
coin. The crucial fact that makes this problem easy is that, although
the involved binomials are by themselves huge numbers (with $\cO(N \ln
N)$ digits), the ratio $p_{n,m}$ is just the simple rational function
$\frac{n}{n+m}$, and various performing \emph{Buffon machines}
\cite{flajBuffon, KY76} can simulate this coin. So, the classical
recursive method has a `good' linear complexity.

The use of the Boltzmann method would go as follows. Consider random
walks of length $N=n+m$, not constrained to the final position, with
i.i.d.\ steps going east or north with probabilities $p$ and
$1-p$. These walks are trivially generated in linear time, and reach
$(n',N-n')$ with probability $\binom{N}{n'} p^{n'}
(1-p)^{N-n'}$. Thus $n'$ is a random variable, centered around $pN$.
However, even using a biased coin at the optimal value for $p$ (and 
neglecting 
the bit complexity of producing this biased coin), 
the probability that $n'=n$ is
only of the order of $N^{-1/2}$,
thus we need to perform on average $N^{1/2}$ independent runs of the
algorithm,
and we have a `bad' overall average complexity $\tT(N^{3/2})$.

Now let us move on to an apparently similar structure: the partitions of
$n+m$ elements
into $n$ non-empty parts. These structures are counted by the
\emph{Stirling numbers of the second kind}, $\stisk{n+m}{n}$
\cite[chapt.~5]{comtet}, and satisfy the linear recurrence relation
\be
\stisk{n+m}{n} = \stisk{n+m-1}{n-1} 
+ n \; \stisk{n+m-1}{n}
\ef.
\label{eq.stirec}
\ee
We stress again that this recursion has an algorithmic couterpart: the
element $n+m$ can either be a singleton (first summand), or can be
inserted in one of the $n$ previous blocks (second summand). If we had
a biased coin of parameter $p_{n,m} = \stisk{n+m-1}{n-1} /
\stisk{n+m}{n}$, and could sample from the ensembles of size up to
$n+m-1$, we could grow our partition by tossing our coin, and, if the
second summand is selected, toss a further integer uniformly in
$\{1,\ldots,n\}$, for choosing the block receiving the new element
(this is done with small complexity $\tT(\ln n)$).
We thus have a formula for average and worst-case complexities
completely analogous to (\ref{eqs.complerec}),
and yet again the whole complexity estimate is pushed towards the
determination of the complexity for the biased coin, 
$T^{\rm coin}_{n,m}$.

Now, despite the apparent similarity of the underlying
recursions (\ref{eq.binoms}) and (\ref{eq.stirec}), in this case there
is no simple formula for $p_{n,m}$. The recursive method would have as
only resort a painful preprocessing of the values $\stisk{n'+m'}{n'}$
for all $n' \leq n$, $m' \leq m$, which is expensive, namely $\tT(N^3
\ln N)$, in terms of both time and space complexities.  One could
reach $\tT(N \ln N)$ space complexity, by recalculating the exact
Stirling tables at all rounds, in small congruence classes, and then
using the chinese remainder theorem, at a price of a $\tT(N^4 \ln N)$
time complexity.

On the other side, the Boltzmann method works along the same lines as
for random walks, thus within linear space, and a time complexity
$\tT(N^{3/2})$ \cite{BasNic}.  
This can be seen, e.g., from the
simple generating function in which we do not fix the number of elements,
but only the number of parts
\begin{align}
\sum_{m \geq 0} \stisk{n+m}{n} x^m
&=
\prod_{y=1}^n \frac{1}{1-x y}
\ef;
&
\stisk{n+m}{n} 
& =
\oint
\frac{\rd z}{2 \pi i z}
\frac{z^{-m}}{\prod_{y=1}^n (1-z y)}
\ef.
\label{eq.68465876}
\end{align}
Here we made use of the Cauchy residue theorem,
and obtained a prototype example of \emph{saddle-point integral} \cite[ch.~VIII]{flaj}.
Note how (\ref{eq.68465876}) agrees with (\ref{eq.stirec}),
as $\oint \frac{\rd z}{2 \pi i z} \big( 1 - zn
- (1-zn) \big) \frac{z^{-m}}{\prod_{y=1}^n (1-z y)} =0$.

These partitions are in bijection with certain rectangular
$(n+m)\times n$ tableaux \cite{BasNic}, whose profile is described by
a sequence of $n$ independent geometric variables $c_y$, with average
$xy$, and total sum $\sum_y c_y=m$. Tableaux with a given profile are
easily uniformly sampled.  The sum over $m$ makes these variables
independent, thus providing with a simple efficient sampling, at the
price of having at most a $\tT(N^{-1/2})$ acceptance probability, a
quantity maximised when $x$ is the unique solution in $[0,n^{-1}]$ of
$x \frac{\rd}{\rd x} \ln \left( \prod_{y=1}^n \frac{1}{1-x y} \right)
= m$, that for large $N$ leads to the transcendental equation
\cite{usMinAuto}
%
\be
\frac{m+n}{n}
=
\frac{-\ln(1-nx)}{nx}
\ef.
\label{eq.sp1}
\ee
Here comes our crucial observation: the saddle-point formula
(\ref{eq.68465876}), besides being at the heart of the Boltzmann
method for this problem, can also efficiently provide good (and
automatisable) \emph{estimates} for our biased coins $p_{n,m}$, the
missing ingredient in the recursive algorithm. The complex-analysis
justification of this claim is well known (see e.g.\ 
\cite[secs.~VIII.2, .3]{flaj}. What is less known is that, with some
extra work (still automatisable),
it is possible to convert these estimates into rigorous upper and
lower bounds.  Better and better estimates will be more and more
computationally expensive, but, for most of our coin tossings, we will
not need a high precision (knowing $d$ binary digits of $p_{n,m}$ is
enough for a fraction $1-2^{-d}$ of the recursive steps).

In our example
we have
\be
1-p_{n,m}
=
\fracdsline{
\oint
\frac{\rd x}{2 \pi i x}
\, (xn) \,
\frac{x^{-m}}{\prod_{y=1}^n (1-x y)}
}
{
\oint
\frac{\rd x}{2 \pi i x}
\, \frac{x^{-m}}{\prod_{y=1}^n (1-x y)}
}
\ef,
\label{eq.ppnm1}
\ee
a quantity which is approximatively given by $x_* n$, where $x_*$ is
the position of the saddle point (\ref{eq.sp1}).

There exists also an alternate saddle-point expression for Stirling
numbers of the second kind. As we deal with ``unlabeled sets of
non-empty sets'', we also have
\be
\stisk{n+m}{n}
=
\frac{(n+m)!}{n!}
\oint
\frac{\rd z}{2 \pi i z}
\frac{(e^z-1)^n}{z^{n+m}}
\label{eq.SPI2}
\ef.
\ee
The position of the saddle point satisfies
\be
\frac{m+n}{n}
=
\frac{z}{1-e^{-z}}
\ef.
\label{eq.sp2}
\ee
(This is the same equation as (\ref{eq.sp1}), if we identify $1-e^{-z}=nx$).

In this case the recursion (\ref{eq.stirec}) is a bit more hidden. We
should use the fact that $\oint \frac{\rd z}{2 \pi i} \left(
\frac{\rd}{\rd z} f(z) \right) = 0$, to get an equivalence with the
relation
\be
0
=
\frac{(n+m)!}{n!}
\oint
\frac{\rd z}{2 \pi i}
\frac{\rd}{\rd z} 
\left(
\frac{(e^z-1)^n}{z^{n+m}}
\right)
\ee
In this framework we find an expression for $p_{n,m}$, alternate 
w.r.t.\ (\ref{eq.ppnm1})
\be
1-p_{n,m}
=
\fracdsline{
\oint
\frac{\rd z}{2 \pi i z}
\, \frac{n\,z}{n+m} \,
\frac{(e^z-1)^n}{z^{n+m}}
}
{
\oint
\frac{\rd z}{2 \pi i z}
\, 
\frac{(e^z-1)^n}{z^{n+m}}
}
\ef,
\label{eq.ppnm2}
\ee
a quantity which is approximatively $\frac{n z_*}{n+m}$, where $z_*$ is
the position of the saddle point (\ref{eq.sp2}).

We have thus arrived at the intuition that, through the idea of
saddle-point estimates, we can improve the recursive
method. In order to make this precise, we need to address three issues:\\
\myitem{i} We need to translate equations like (\ref{eq.ppnm2}) into exact
  bounds, of the form
  $\xi - \e \leq p \leq \xi + \e$, where the functions $\xi=\xi(n,m,z_*)$ and
  $\e=\e(n,m)$ are sufficiently explicit to admit fast bit-complexity
  evaluations at the required $o(\e)$ precision (roughly speaking,
  these functions can be defined in terms of special functions such as exponentials or
  logarithms, but not through transcendental equations).\\
\myitem{ii} In most of the interesting cases, including (\ref{eq.sp2}), the
  expression $z_*=z_*(n,m)$ is the solution of a transcendental
  equation, so we need an efficient numerical approximation method,
  and we must control the propagation of the error in
  $\xi(n,m,z_*(n,m))$.\\
\myitem{iii} 
We branch ``left'' or ``right'' if the random value $x \in [0,1]$ is
$x<\xi-\e$ or $x>\xi+\e$. We need to resolve the case $x \in [\xi-\e,
  \xi+\e]$.  This may be done through a tighter bound, that uses one
more term in the Taylor or Euler-Maclaurin expansions pertinent to the
saddle point analysis, or even through a standard step of the
recursive method, with the exact construction of the branching
probabilities, when $2\e$ is small enough so that the associated
average complexity is negligible.

In the following sections we outline a general strategy to address
these issues, for large families of specifiable combinatorial
structures, and describe sufficient conditions for our strategy to
apply, easy to verify on any given problem. All along the paper, we
illustrate this automatised construction on the example of
partitions discussed above.

\section{Linear-time recursive method with oracles}
\label{sec.orac}

In order to pursue the idea above, it is convenient to separate the
study into two parts. In this section, we show that, given a hierarchy
of \emph{oracles} for these bounds, assumed to cost a certain
complexity, the average complexity of the recursive method would be
quasi-linear. Next, in Section~\ref{sec.oracc} we show how these
oracles are implemented, with the announced
complexity, from the saddle point expressions.

We assume to have statistical ensembles of combinatorial structures
$X$, with $d$ size parameters, $X \in \cX_{\bn}$, $\bn = (n_1, \ldots,
n_d) \in \bbN^d$, and we want to sample from some measure, that could
possibly depend on further real-positive weight parameters (that
will be considered as fixed, and whose dependence is left
implicit). We call $N=\sum_j n_j$.

We also assume to have some generating functions $Z_{\bn}$ associated
to these measures, for which we know saddle-point expressions.  We
suppose to have a recurrence relation, of the form
\be
Z_{\bn} 
= \sum_{j=1}^{k+1}
c_j(\bn)\, Z_{\bn - \bv_j}
\ee
where the $\bv_j$'s are vectors in $\bbN^d \smallsetminus {\bm 0}$,
and the $c_j(\bn)$'s can be computed easily. We assume that the
relation above is associated to a recursive construction of the
objects: one can sample from the ensemble $\cX_{\bn}$, by choosing $1
\leq j \leq k+1$ with probability $\xi_j-\xi_{j-1}$, where,
%
$
\xi_j = 
Z_{\bn}^{-1}
\sum_{i=1}^j
c_j(\bn) Z_{\bn - \bv_i}
$,
%
then sampling from $\cX_{\bn-\bv_j}$, and finally performing a further
algorithmic step for growing the structure, of complexity
$T^{\rm step}_{\bn}\leq P \cdot (\ln N)^p$
(as discussed above, for the examples of directed walks and partitions
of $n$ elements into $k$ blocks we have $p=0$ and $p=1$ respectively).
Thus, we have an `intrinsic complexity' of the recursive method,
$T_{\rm intr}(N) \leq \sum_M P \cdot (\ln M)^p \leq P N (\ln N)^p$,
that would be the complexity in the idealised paradigm in which the
oracle and the sampling of random numbers for the branching procedure
have zero cost. This complexity summand is inherent to the recursive
method, and ineliminable (unless one changes completely the algorithm,
and e.g.\ finds a more efficient construction). As our main point here
is the optimisation of the branching procedure, we will not address
the issue of optimising $T_{\rm intr}(N)$.

Let us denote $\cN_{\rm easy}(N) \subset \bbN^d$ the set of values
$\bn$ for which the sampling is performed more efficiently with some
different method, with complexity $T_{\rm easy}(\bn)$, such that
$T_{\rm easy}(\bn) = o(N)$ for all $\bn \in \cN_{\rm easy}(N)$.
A simple general choice is $\cN_{\rm easy} = \{ \bn \;|\; \sum_j n_j
\leq N_0 \}$, where $N_0 = o(N^{1/\alpha})$, and $\alpha$ is the
smallest complexity among the ordinary recursive and Boltzmann
algorithms.  Note that, as a result, the recursive construction always
halts at sizes $\gg 1$, and our uncertainty on the thresholds $\xi$
will be $\ll 1$ at all steps.
For some special problems, $\cN_{\rm easy}$ could be larger. For
example, it could include certain extreme ranges of parameters, $n_j/N
= o(1)$ for certain $j$, even for large
\label{pref.footNeasy} 
$N$.\;\footnote{This is the case for Stirling numbers of the second
  kind. If the number of parts $k$ is sub-linear w.r.t.\ the number of
  elements $n$, we can try to randomly colour our elements, with
  labels from $1$ to $k$ and thus with complexity $n \ln k$, and
  reject the result if any colour is not used, event of probability
  bounded by $k \exp(-n/k)$ and thus of order $1$ if $k \ln k \ll
  n$. In the opposite regime, the number of parts being $n-k$, with
  $k$ sub-linear, we can randomly sample $k$ edges of $\mathcal{K}_n$,
  use the connected components as parts of the partition, and accept
  the resulting configurations: (i) never, if the graph contains any
  loop; (ii) otherwise, with probability $p = \prod_j
  (j^{j-2})^{-C_j}$ if we have $C_j$ tree components of size $j$. Note
  that the factors for $j=1,2$ are just 1. The probability of having
  any loop at all is bounded by classical results on Erd\H{o}s-R\'enyi
  graphs, while the expected $C_j$ for $j \geq 3$ is of order
  $k^{j-1}/n^{j-2}$, thus $o(1)$ as long as $k \ll \sqrt{n}$.}


We assume that, for some integer $s_{\rm Max}$, we have a hierarchy of
estimates, of the form
$\xi_j^{\ell, s} \leq \xi_j\leq \xi_j^{u, s}$ with $\xi_j^{u, s} -
\xi_j^{\ell, s}\leq g_s N^{-s}$, for each level $s < s_{\rm
  Max}$. These intervals of uncertainty may overlap, e.g. it may be
that $\xi_{j}^{u, s} > \xi_{j+1}^{\ell, s}$, although we know that, by
construction, $\xi_j < \xi_{j+1}$.

Our algorithm
goes as follows: when at size $M$, sample $x \in [0,1]$, and evaluate
$\xi_j^{\ell, 1}$ and $\xi_j^{u, 1}$ for all $1 \leq j \leq k+1$. If
$j$ is determined univocally, i.e.\ $\xi^{u,1}_{j-1} < x <
\xi^{\ell,1}_{j}$ for some $j$, (this happens with probability at least $1-k
g_1/M$), we go on along the appropriate branch, as in an ordinary
recursive algorithm. Otherwise, we need to consider the tighter bound
at $s=2$, and so on. The probability of having to consider a bound of
level $s \geq 2$ is at most $k g_{s-1} M^{-s+1}$.  We set $g_0=1$, in
order to make this formula valid at all $s$.  If not even the last
bound at $s=s_{\rm Max}-1$ is tight enough, we perform an ordinary,
exact recursive method.

Let $T_s(M)$ be an upper bound to the complexity for the evaluation of
the level-$s$ bounds,
and $T_{\rm exact}(M)$ be a bound to the exact recursive method.  With
respect to the idealised recursive method, with zero-cost oracles, the
average complexity has extra terms, of order $\sum_{M=N_0}^N g_{s}
M^{-s} \, T_{s+1}(M)$, from the use of the level-$s$ bound, and of
order $\sum_{M=N_0}^N g_{s_{\rm Max}} M^{-s_{\rm Max}} T_{\rm
  exact}(M)$, from the use of our `last resort' exact method.
%
Under moderate assumptions on our complexities, $T_{s}(M) \leq Q_s
\cdot (\ln M)^{q_s}$ and $T_{\rm exact}(M) \leq M^{\gamma}$, and
choosing $s_{\rm Max} > \gamma +1$, the overall cost is dominated
either by the intrinsic complexity, $T_{\rm intr}(N) = \tT(N (\ln
N)^p)$, or by the determination of the level-1 bounds, which takes
$\tT(N (\ln N)^{q_1})$. In some cases, one can ensure that
the latter logarithmic prefactor $(\ln N)^{q_1}$ does not exceed the
intrinsic one $(\ln N)^p$, by producing the bits of the level-1 bounds
as long as they are needed, and performing a realistic analysis at the
level of bit complexity.  We do not do this here,\footnote{See
  Appendix~\ref{app.binary} for a partial discussion.} and
we just summarise the result of the analysis.
\begin{proposition}
\label{prop.orac}
Consider a recursive algorithm, with complexities: \\
\myitem{1} $R$ for extracting a random bit;\\
\myitem{2} $T_{s}(M) \leq Q_s \cdot (\ln M)^{q_s}$ for producing
  $\lrceil{\log_2 M^s/g_s}$ digits of the bounds $\xi_j^{\ell, s}$
  and $\xi_j^{u, s}$;\\
\myitem{3} $T_{\rm exact}(M) \leq M^{\gamma}$ for performing an
  ordinary exact recursive step;\\
\myitem{4} $C_d$ for querying the $d$-th digit of $\xi_j^{\ell,1}$, $\xi_j^{u,1}$, 
  if the first $d-1$ ones are known, with
  $C_d \leq H d^h e^{\eta d}$.\\
If $\gamma < s_{\rm Max}-1$, and $\eta < \ln 2$, then the
algorithm runs with average bit complexity bounded by
\begin{align}
T 
& \leq
T_{\rm intr}(N) +
K\; N
+ o(N)
\ef;
&
K
&=
3kR + k H h! (1-e^{\eta -\ln 2})^{-h-1}
\ef.
\end{align}
If the hypothesis (4) does not hold, we still have
\be
T 
\leq
T_{\rm intr}(N) +
\big( 3kR + Q_1 (\ln N)^{q_1} \big) N
+ o(N)
\ef.
\ee
\end{proposition}

\section{Construction of the oracles}
\label{sec.oracc}

In this section we explain how one can systematically construct a
hierarchy of oracles satisfying the complexity constraints of
Proposition \ref{prop.orac}, when the unnormalised measures $Z_\bn$
are expressed through a saddle-point integral, in which the integrand
has a sufficiently simple form.  This is done in subsection
\ref{sec.HieraBou}.\,\footnote{An extension to a larger class of
  integrands is discussed in Appendix~\ref{sec.eulml}.}  In order to
do so, we need some preliminary technical results, discussed in
subsections \ref{sec.Sxy} and \ref{sec.polybounds}.  A subtle
issue on how to determine efficiently the position of the saddle point
is discussed in subsection~\ref{sec.follow}.

\subsection{Formal solution of $S(x(y))=y^2$}
\label{sec.Sxy}

While the systematic expansion in $n^{-1}$ of saddle point integrals
can be performed in several equivalent ways, and among them through
the brute-force Taylor expansion of the non-quadratic part of the
action, the resulting bounds are more or less performing, depending on
the used construction, and some new special tricks come into play.

One of them is the solution of the equation $S(x(y))=y^2$, given that
$x(y)=y+a_2 y^2+a_3 y^3+\ldots$\ and $S(x)=x^2+b_3 x^3+b_4
x^4+\ldots$\; ($x(y)$ and $S(x)$ are formal power series).  There
exist two versions of the problem: finding the appropriate series $a$,
given $b$, or finding $b$ given $a$.  We thus need to solve, for all
$k \geq 3$, $C_k(a,b) := [y^k]S(x(y))=0$ (lower degrees are matched
automatically).

The relevant observation is that $C_k(a,b)=2 a_{k-1}+b_k+C'_k(a,b)$,
where $C'_k$ is a polynomial depending only on the indeterminates
$a_2,\ldots,a_{k-2}$ and $b_3,\ldots,b_{k-1}$. Thus the system of
equations is triangular, for both versions of the problem.

The first few terms for $b(a)$ read
\begin{align}
\begin{split}
b_3 
&
= \makebox[0pt][l]{$-2 \, a_{2}\ef;$}\rule{150pt}{0pt}
\makebox[0pt][r]{$b_4$} = 5 \, a_{2}^2 - 2 \, a_{3} 
\ef;
\\
b_5 
&
= \makebox[0pt][l]{$-14 \, a_{2}^3 + 12 \, a_{2} a_{3} - 2 \, a_{4}\ef;$}\rule{150pt}{0pt}
\makebox[0pt][r]{$b_6$} = 42 \, a_{2}^4 - 56 \, a_{2}^2 a_{3} + 7 \, a_{3}^2 + 14 \, a_{2} a_{4} - 2
\, a_{5} 
\ef;
\\
b_7 
&
= -132 \, a_{2}^5 + 240 \, a_{2}^3 a_{3} - 72 \, a_{2} a_{3}^2 - 72
\, a_{2}^2 a_{4} 
+ 16 \, a_{3} a_{4} + 16 \, a_{2} a_{5} - 2 \, a_{6} 
\ef.
\end{split}
\intertext{The solution for $a(b)$ is best visualised separating even and odd
coefficients. The first terms are}
\begin{split}
2a_2 &= 
\makebox[0pt][l]{$-b_{3} \ef;$}\rule{150pt}{0pt}
\makebox[0pt][r]{$2a_4$} = -2 \,b_{3}^3 + 3 \,b_{3} b_{4} - b_{5}\ef; \\
2a_6 &= -7 \,b_{3}^5 + 20 \,b_{3}^3 b_{4} - 10 \,b_{3} b_{4}^2 - 10
\,b_{3}^2 b_{5} 
+ 4 \,b_{4} b_{5} + 4 \,b_{3} b_{6} - b_{7}
\ef;
\end{split}
\intertext{and}
\begin{split}
2^{3} a_3 &= \makebox[0pt][l]{$5 \,b_{3}^2 - 4 \,b_{4}\ef;$}\rule{150pt}{0pt}
\makebox[0pt][r]{$2^{7} a_5$} = 231 \,b_{3}^4 - 504 \,b_{3}^2 b_{4} +
112 \,b_{4}^2 + 224 \,b_{3} b_{5} - 64 \,b_{6} 
\ef;
\\
2^{11} a_7 &= 14586 \,b_{3}^6 - 51480 \,b_{3}^4 b_{4} + 41184 \,b_{3}^2
b_{4}^2 - 4224 \,b_{4}^3 
+ 27456 \,b_{3}^3 b_{5} - 25344 \,b_{3} b_{4} b_{5} 
\\& \quad
+ 2304 \,b_{5}^2 - 12672 \,b_{3}^2 b_{6} + 4608 \,b_{4} b_{6} + 4608 \,b_{3} b_{7} -
1024 \,b_{8}
\ef.
\end{split}
\end{align}
Note that, in our applications, we will only need the solution $a(b)$
up to order $s_{\rm Max}$, thus, for every problem, where 
$s_{\rm Max}$ is fixed and determined by the complexity of the ordinary
recursive step, this is a fixed $\cO(1)$ preprocessing.

\subsection{Polynomial bounds to analytic functions}
\label{sec.polybounds}

We introduce here a convenient notation for calculating error bounds
in the complex plane, that generalizes the standard ``$\pm$'' notation
for error propagation from elementary statistics on~$\bbR$.  For $A,B
\in \bbC$, denote as customary $f(A)\equiv\{f(a)\}_A$, $AB=\{ab\}_{a
  \in A, b \in B}$ and $A+B=\{a+b\}_{a \in A, b \in B}$.  For $a \in
\bbC$ and $b \in \bbR^+$, let $a \ptil b$ denote the disk in $\bbC$ of
center $a$ and radius $b$.
This notation has several nice properties, such as
\begin{align}
\label{eq.ptilPlus}
(a \ptil b) + (c \ptil d) 
&= (a+c) \ptil (b+d)
\ef;
\\
c(a \ptil b) 
&= ca \ptil |c|b
\ef;
\end{align}
and, when $f(z)$ is analytic, as an analytic function on $D$ always takes
its maximum on $\partial D$, $f(a \ptil b) \peq f(a) \ptil b'$, with
$b'=\max_{\theta} |f(a+b e^{i \theta})-f(a)|$.

Among the corollaries of this fact, we have for any real positive $b$ 
\be
\label{eq.57686456a}
\exp(\ptil b) \peq 1 \ptil (e^b-1)
\ef,
\ee
and for real positive values $a$, $b$, $c$, $d$ such that
$a>b$, $c>d$ 
\be
\label{eq.57686456}
\frac{a \ptil b}{c \ptil d}
\peq
\frac{1}{c^2-d^2}
\big( (ac+bd)
\ptil (ad+bc) \big)
\ef.
\ee
We also have, for $P(z)=p_1 z+p_2 z^2+\ldots+p_d z^d$ a polynomial,
\be
\label{eq.expPz}
e^{P(z)} \in
e^{p_1 z} e^{\ptil (|p_2 z^2| 
+ \cdots + |p_d z^d|)}
\subseteq
e^{p_1 z} \bigg(
1 \ptil
|z|^2 \frac{e^{|p_2| \eta^2 
+ \cdots + |p_d| \eta^d}-1
}{\eta^2}
\bigg)
\qquad
|z| \leq \eta
\ef.
\ee
We need a similar result for generic functions.
Consider the function $f(z)=f_0 + f_1 z + f_2 z^2 + \cdots$, analytic
and with radius of convergence $\rho$, and call $f^{[k]}(z) = f_0 +
f_1 z + f_2 z^2 + \cdots + f_{k-1} z^{k-1}$.  For $\eta < \rho$, we
want to determine a function $r(\eta)$ such that $f(z) \in f^{[k]}(z)
\ptil r(\eta) |z|^k$.  Assume that all coefficients $f_j$ are real
positive, for $j \geq k$. Then the maximum on the disk $D$ of radius
$\eta$ is realised for $z=+\eta$, and we have
\be
\begin{split}
|f(z) - f^{[k]}(z)|
&=
\sum_{j \geq k} f_j |z|^j
\leq |z|^k \sum_{j \geq k} f_j |\eta|^{j-k}
=
|z|^k \frac{f(\eta)-f^{[k]}(\eta)}{\eta^k}
\ef;
\end{split}
\ee
so that we can state
\be
f(z) \in f^{[k]}(z) \ptil
|z|^k
\frac{f(\eta)-f^{[k]}(\eta)}{\eta^k}
\ef.
\ee
If $f_j$'s are all negative, or have alternating sign,
it suffices to take
$|f^{[k]}(\eta)-f(\eta)|$
or
$|f^{[k]}(-\eta)-f(-\eta)|$.  If we have an explicit decomposition
$f(z)=\sum_{\sigma, \tau = \pm} f_{\sigma \tau}(z)$, where, for $j
\geq k$, $[z_j] f_{\sigma \tau}(z)$ has sign $\sigma$ or $\tau$
depending if $j$ is even or odd, we can use the previous estimates
separately on the four terms, and recombine them using
(\ref{eq.ptilPlus}).  Let us call $\cF_{\sigma,\tau}^k$ the convex
cone of analytic functions $f$, non-singular in $z=0$, such that
even/odd coefficients $f_j$ with $j \geq k$ have sign $\sigma$ and
$\tau$, respectively. We say that
$\{f_{\sigma,\tau}(x)\}_{\sigma,\tau=\pm}$ is a
\emph{$k$--sign-decomposition} of $f(z)$, if
$f(x)=\sum_{\sigma,\tau=\pm} f_{\sigma,\tau}(x)$ and
$f_{\sigma,\tau}(x) \in \cF_{\sigma,\tau}^k$.
Of course, if $P$ is a polynomial with real coefficients, of degree
$d$, and we have a $k$--sign-decomposition of $f$, we have a
straightforward $k$--sign-decomposition of $f+P$ (if $d>k$, just
attribute positive and negative coefficients of $P$ to $f_{++}$ and
$f_{--}$, respectively),
a fact that we use later on.


Sign-decompositions may look abstract, but are in fact easily obtained
in various concrete circumstances. Let us illustrate this within our
case example, i.e.\ for the action $S(z)$ (logarithm of the integrand)
in equation (\ref{eq.SPI2}).  Let us parametrise this function
according to the position $\z$ of the saddle point. We thus have,
using (\ref{eq.sp2}),
$\frac{m+n+1}{n}=\frac{\z}{1-e^{-\z}}$ (so that $\z \in \bbR^+$), and,
up to a rescaling,
\be
S_\z(x)=
(1 - e^{-\z}) \ln(e^{\z + x} - 1) -\z \ln(\z + x)
\ef.
\label{eq.7863545}
\ee 
We want to present a sign-decomposition (in the variable $x$) that
holds \emph{simultaneously for all values of $\z$} in the range. It is
not evident \emph{a priori} that this is possible. But in fact the two
summands of (\ref{eq.7863545}) are in $\cF_{+-}^1$ and $\cF_{-+}^1$,
respectively. This is obvious for the second one. For the first one,
use the striking fact
\be
\ln \frac{e^x-y}{1-y}
=
\frac{x}{1-y}
+y
\sum_{n,k}
\frac{(-1)^{n-1}}{n! (1-y)^n}
T_{n,k} x^n y^k
\label{eq.EulNums}
\ee
where the coefficients $T_{n,k}$ are the \emph{Eulerian numbers}
(number of permutations of $n+1$ objects with $k$ rises)
\cite[sect.~6.5]{comtet}, and in particular they are all positive
integers.  Our first summand is related to the expression above,
identifying $y=e^{-\z} \in [0,1]$, thus it is in $\cF_{-+}^1$.

\subsection{The hierarchy of saddle-point bounds}
\label{sec.HieraBou}

Suppose you want to evaluate a hierarchy of bounds to the quantity,
analogous to equation (\ref{eq.ppnm1}),
\be
\xi_{n}
=
\fracdsline{
\oint
\frac{\rd z}{2 \pi i}
A(z) \,
\exp(n S(z))
}
{
\oint
\frac{\rd z}{2 \pi i}
B(z) \,
\exp(n S(z))
}
\ef.
\label{eq.45179176}
\ee
Suppose that $A$ and $B$ are polynomials, that the dominant
saddle point $z_*$ is isolated, on the positive real axis, that
$S''(z_*)>0$ (so that the steepest-descent Cauchy contour is vertical
near to the saddle point), and that we have an allowed topology of
contour around the origin,
of finite length.  Essentially all of these requirements can be
relaxed, at the price of making the discussion more
convoluted.\,\footnote{For example, the treatment is easily extended
  to the case of $A$ and $B$ with a sign-decomposition, each summand
  of the decomposition having no singularities in a neighbourhood of
  the saddle point, as these can be rephrased into polynomials, up to
  bounds of the form $\ptil r(\eta) |z-z_*|^k$ for arbitrary large
  $k$.}
In particular, the requirement of singularity on $\bbR^+$ is not
really restrictive, as in fact it is essentially implied by the
requirement of having a recursive description with positive
coefficients.
Up to a rescaling of the variables $z$ and $n$, we can set
$z_*=S''(z_*)/2=1$.

We want to determine a finite sequence of complex numbers $a_s$, and
functions $r_s(\eta): \bbR^+ \to \bbR^+$ such that, for all $s \leq
s_{\rm Max}$, setting $x(y)=y+a_2 y^2+ \cdots + a_s y^s$,
\be
|S(1+ix(y)) + y^2|
\leq r_{s}(\eta) |y|^{s+1}
\qquad
\forall \; |y| \leq \eta
\ef;
\ee
i.e.,
$S(1+ix(y)) \in -y^2 \ptil r_{s}(\eta) |y|^{s+1}$.

From Section \ref{sec.Sxy} we know explicitly the unique candidate
series $a_i$, in terms of the first $s_{\rm Max}$ derivatives of
$S(z)$ at the saddle point.  We need $r_s(\eta)< \infty$ when $\eta$
is large enough for our purposes. As we will see, since we assumed
that $z_*$ is isolated, this will always be the case for $n$ large
enough.  However, the existence of $r_s(\eta)$ is not by itself
sufficient. As the bounds are expressed in terms of this function, we
need it to be \emph{computable}. At the light of the results of
Section \ref{sec.polybounds}, we have an automatised construction if
we know an explicit sign-decomposition of $S(1+x)$.

Let us concentrate on the numerator of (\ref{eq.45179176}).  Fix
$\eta$ such that $r_s(\eta)<\infty$, and call $x_{\pm} = x(\pm
\eta)$. 

Divide the contour path $\gamma$ into the path 
$\gamma_{\rm gauss}$ image of $[-\eta,\eta]$ w.r.t.\ $z_*+x(y)$, and a
path $\gamma_{\rm rest}$ from $x_+$ to $x_-$ encircling the origin
from the left. We thus have
\be
\begin{split}
&
\oint_{\gamma}
\frac{\rd z}{2 \pi i}
A(z) \,
\exp(n S(z))
=
\left(
\oint_{\gamma_{\rm gauss}}
+
\oint_{\gamma_{\rm rest}}
\right)
\frac{\rd z}{2 \pi i}
A(z) \,
\exp(n S(z))
\end{split}
\ee
and we will concentrate on the first summand (we easily bound the
second summand at the end). Write $z=z_*+ix(y)$. Make a change of
variables from $z$ to $y$, with Jacobian $J(y) = i \rd x(y)/\rd y = i (1+2 a_2
y + \cdots + s a_s y^{s-1})$, to obtain an integral proportional to
\be
\int_{-\eta}^{\eta}
\!\!\!
\rd y
\,
J(y)
\,
A(z_*+ix(y))
\,
e^{n (-y^2 \ptil r_{s}(\eta) |y|^{s+1})}
\ef.
\ee
The quantity $J(y) A(z_*+ix(y))$ is a certain polynomial 
$P(y)=p_0 + p_1 y + \cdots + p_c y^c$, 
where the coefficients $p_i$ depend on $z_*$ and the $a_j$'s. We can
use (\ref{eq.57686456a}) on the remainder term in the exponential.
So, calling
$R= (e^{n r(\eta) \eta^{s+1}}-1)\eta^{-(s+1)}$,
the integral above is inside the disk
\be
\sum_{j=0}^c p_j
\int_{-\eta}^{\eta}
\!\!\!
\rd y
\, y^j
(1 \ptil R |y|^{s+1})
e^{-n y^2}
\ef.
\ee
We sum and subtract the integral on the intervals 
$(-\infty,-\eta]$ and $[\eta,+\infty)$ (we consider the subtracted
quantities together with the integral $\oint_{\gamma_{\rm rest}}$).
We are thus led to the study of integrals of the form
\begin{align}
I^-_j(n)
&=
\int_{-\infty}^{\infty}
\!\!\!
\rd y
\, y^j
e^{-n y^2}
\ef;
&
I^+_j(n)
&=
\int_{-\infty}^{\infty}
\!\!\!
\rd y
\, |y|^j
e^{-n y^2}
\ef;
\end{align}
which give the well-known formula
\be
I^{\epsilon}_j(n)
=
\frac{1+\epsilon^j}{2}
n^{-\frac{j+1}{2}} 
\Gamma\Big(\frac{j+1}{2}\Big)
\ef.
\ee
In most cases one can easily bound the portion of the integral
associated to $\gamma_{\rm rest}$ by some quantity of the form
$T(n,\eta) = T z_*^\tau \exp(-n(\eta^2-r(\eta) \eta^{s+1}))$, for some
finite $\tau$. This is discussed, e.g., in \cite[sec.~VIII.3]{flaj},
and, for the most frequent problems, this issue has already been
solved explicitly in the literature concerning asymptotic
enumeration.\,\footnote{A simple criterium, applicable in many cases
  including our case example, is given in
  Appendix~\ref{sec.tails}.}  As the function $r_s(\eta)$ is smooth
and finite near $\eta=0$, the function $\eta^2-r_s(\eta) \eta^{s+1}$
has a positive maximum for some $\eta>0$, which is a locally optimal
value for our bounds of these terms.

So we have
\be
\xi_n = 
\frac{A \ptil (
\delta A
+ A_{\rm rest})}
{B \ptil (
\delta B
+ B_{\rm rest})}
\ee
with
\begin{align}
A
&= 
\sum_{j=0}^{\lrceil{c}} p_{2j} 
n^{-j-\frac{1}{2}} 
\Gamma\Big(j+\frac{1}{2}\Big)
\ef;
&
\delta A
&=
R
\sum_{j=0}^c p_{j} 
n^{-\frac{j+s+3}{2}}
\Gamma\Big(\frac{j+s+3}{2}\Big)
\ef;
\end{align}
and 
$
A_{\rm rest}
=
T(n,\eta) 
$.
We proceed similarly for $B$, and simplify the ratio
using (\ref{eq.57686456}), to produce upper and lower bounds.


\subsection{Following the saddle point}
\label{sec.follow}

As we have seen, we are in general able to construct analytic
expressions for our bounds, that depend on the rational parameters
$\a_j = n_j/N$ both directly and through the location of the saddle
point $z_*(\ba)$. The equation that determines $z_*$ from $\ba$,
however, is often transcendental. Thus, apparently it should be solved
numerically at each round, at the appropriate precision, this being
computationally expensive.
We can improve on this, by exploiting two facts:
\begin{enumerate}
\item Even if we miss the location of the saddle point `by a tiny
  bit', a less tight version of bounds still exist. In this case one
  could treat the error factor $\exp(N \epsilon\, x(y))$ through equation
  (\ref{eq.expPz}).
  If we have $\epsilon=o(N^{-1})$, essentially nothing happens at the
  level of the first bound, which is the dominant source of
  complexity, as we know from Section \ref{sec.orac}.  More generally,
  $\epsilon=o(N^{-\frac{s+1}{2}})$ suffices to have no effect on the
  level-$s$ bound.
\item The saddle point moves slowly. Namely, if the singularity is
  isolated, the proper root of the
  equation $S'_\bn(z)=0$ has no multiplicity, and its variation is
  linear, $|z_*(\bn -\bv_j) - z_*(\bn)|/|z_*(\bn)| = \cO(N^{-1})$.  If
  we determined $z_*(\bn)$ up to an error $\cO(N^{-\gamma})$, we
  already know $z_*(\bn -\bv_j)$ up to an error
  $\cO(\max(N^{-\gamma},N^{-1})) = \cO(N^{-\min(\gamma,1)})$.  Let us
  then perform just one step of Newton iteration
  \cite[sec.~6]{brunoNewt}. The error is squared,
  i.e.\ $\cO(N^{-2\min(\gamma,1)})$.
  Thus, as $2\min(\gamma,1)=\gamma$ is only solved by $\gamma=0$ and $2$,
  if we find the value $z_*$ for the initial step of our algorithm, at
  precision $\cO(N^{-2})$, we will keep this level of precision at all
  times just by performing a single Newton iteration at each step. When we
  need to use a higher level bound (this happens on average finitely
  many times on the full run), it suffices to perform a few more
  Newton iterations to determine $z_*$ at higher precision. The
  average total number of required Newton iterations is thus $\cO(N)$.
\end{enumerate}

\section{Summary of sufficient conditions for applying our
  method}
\label{sec.summconds}

We now evince, from the construction of the previous sections, a list
of sufficient conditions to be verified on a given combinatorial
problem, for it to be amenable to our method.  These conditions are
essentially analytic, certified by finite expressions, and normally
easily achieved from any given explicit combinatorial specification.
Once the conditions are established, the construction of the algorithm
is automatised.\\
\myitem{1} You need a recursion relation $Z_{\bn} = \sum_{j=1}^{k+1}
  c_j(\bn) Z_{\bn - \bv_j}$, with positive $c_j$'s, that translates
  into an algorithmic recursive construction.\\
\myitem{2} You need to establish a bound on a single `ordinary'
  recursive step, of the form $T_{\rm exact}(M) \leq
  M^{\gamma} (\ln M)^{\gamma'}$. You can then set $s_{\rm Max}$ to the
  smallest integer strictly larger than $\gamma +1$.\\
\myitem{3} You should establish a set $\cN_{\rm easy}(N)$ where the
  generation is sublinear, and the pertinent alternate algorithm. This
  may consist of all $\bn$ such that $\sum_j n_j = o(N^{1/\a})$, with
  notations as on page~\pageref{pref.footNeasy}.\\
\myitem{4} You need a saddle-point expression for the partition functions,
  $Z_\bn = \oint \frac{\rd z}{2 \pi i z} A(z) \exp(S_\bn(z))$. You
  shall determine the associated saddle-point equation, the
  appropriated topology of the contour, and a bound on the tail terms.\\
\myitem{5} You must write $S_\bn(z)$ as a sum, where each summand depends
  from a unique size variable $n_j$, and produce a sign-decomposition
  for each of these summands.\\
\rule{0pt}{14pt}Many examples of specifiable combinatorial structures
arising in the literature (in particular, in the extensive compendium
of \cite{flaj}) are accessible to the criteria above,
and their systematic analysis
opens up a wide range of applications.

\newpage

\subsection*{Acknowledgements}

We thank Julien David and Bruno Salvy for useful discussions.
This work is supported by the ANR 2010 Magnum BLAN-0204.


\newpage

                        \appendix

\section{Some aspects of bit complexity for the branching procedure}
\label{app.binary}

In Section \ref{sec.orac} we analysed the complexity of the branching
part of the recursive method, under the arithmetic paradigm. This
accounted for attributing a unit cost to the evaluation of functions
up to precision $\tT(N^{-s})$, and thus requiring $\cO(\ln N)$ bits,
and, similarly, allowed to sample ``random real numbers in $[0,1]$'',
and compare them to these functions, again within a unit cost.

Clearly, a more scrupolous analysis of the bit complexity is
mandatory.  It is nowadays a standard result that the evaluation with
$d$ digits of precision of an expression involving certain classes of
elementary functions requires a complexity scaling as $d^{\gamma}$,
where $\gamma$ is a finite exponent depending on the class of
functions entering the expression (see e.g.\
[D.E.~Knuth, {\it The Art of Computer Programming,} Addison-Wesley,
  1998], in particular vol.~2, chapt.~4). We do not enter here in the
details of this wide branch of Theoretical Computer Science.

Instead, the paradigm for the extraction of a ``random real number''
$x \in [0,1]$ is quite easy to describe: we extract the binary digits
of $x$ one at the time, as long as needed.
Let $\xi
\in [0,1]$ be a threshold value, of which we can query the binary digits,
one by one,
with complexity $C_j$ for the $j$-th query, and let $R$ be the complexity
for querying a random bit. The average complexity $T(\xi)$, to determine if $x
\in [0,1]$ is smaller or larger than $\xi$, 
is in fact independent of $\xi$, as the probability of halting at the
$k$-th digit is always $2^{-k}$, and thus we have
$T(\xi) = (R+C_1) + \frac{1}{2} (R+C_2) + \frac{1}{4} (R+C_3) + \ldots = 2R + 
\sum_{i \geq 1} 2^{-i+1} C_i$.

In our situation, the result is only slightly different. We have two
thresholds, $\xi^u$ and $\xi^{\ell}$, and we need to determine if $x
\leq \xi^{\ell}$, $\xi^{\ell} < x \leq \xi^u$ or $x > \xi^u$. We now
know that
the thresholds are dyadic numbers, with $d$ digits of precision, which
differ by $2^{-d}$. In this case the complexity 
is not uniform anymore, and the worst case is
\begin{align}
\xi^{\ell} &= 0.0\underbrace{1\cdots 1}_{d-1}
\ef;
&
\xi^{u}    &= 0.1\underbrace{0\cdots 0}_{d-1}
\ef;
\end{align}
for which we have the slightly modified expression $T(\xi) = (R+2C_1)
+ (R+C_2) + \frac{1}{2} (R+C_3) + \ldots + \frac{1}{2^{d-2}} (R+C_d)
\leq 3R + \sum_{i=1}^{d} 2^{-i+2} C_i$. The best case, obtained for
$\xi^{\ell} = 0.0\cdots 00$ and $\xi^{u} = 0.0\cdots 01$, gives a
quite similar lower bound, $T(\xi) \geq (2-2^{-d})R + \sum_{i=1}^{d}
2^{-i+2} C_i$.

Under the assumption, presented in the hypothesis (4) of Proposition
\ref{prop.orac} and coherent with our discussion on the complexity of
evaluating expressions at a given precision, that $C_d \leq H d^h
e^{\eta d}$, with $\eta < \ln 2$, the sums above converge, and we
obtain the bound $T(\xi) \leq 3R + H h! (1-e^{\eta -\ln 2})^{-h-1}$
appearing in the forementioned proposition.  At this aim it is useful
to perform the approximation $\sum_{j \geq 0} j^h e^{-aj} \leq h!
\sum_{j \geq 0} \binom{j+h}{h} e^{-aj} = h! (1-e^{-a})^{-h-1}$.

\section{Sums of logs and Euler--Maclaurin}
\label{sec.eulml}

Our construction of the bounds presented in Section \ref{sec.oracc},
as a function of $\bn$ but performed once and for all in a
preprocessing phase of constant complexity, assumes that we can
present a sign-decomposition, and the analytic evaluation of the
derivatives of the action, valid for all ranges of $\ba=\bn/N$. A case
in which this is possible is when the action has the form
\be
S_{\bn}(z) = \sum_j n_j S_j(z)
\ef,
\label{eq.decoS}
\ee
and for each $S_j$ a sign-decomposition is produced. As we have seen,
the saddle point integral (\ref{eq.SPI2}) is already in this form. But
this does not necessarily occur in all the problems we aim to
analyse. For example, already in our alternate expression
(\ref{eq.68465876}) we encountered a different situation, as we have
(we change variables from $z$ to $z/n$, and omit an overall
constant)\,\footnote{Note that we already know that, in this case, we
  should perform the integral on a countour with $|z|<1$, so that we
  have no troubles with the radius of convergence of the log.}
\be
S_{(n,m)}
=
-m \ln (z)
-
\sum_{y=1}^n
\ln \Big( 1-\frac{z y}{n} \Big)
\ef.
\label{eq.467654765b}
\ee
Apparently, we need to construct dynamically our bounds for all the
different values of $n$ we encounter, a procedure that would be too
expensive. In fact this computation can be avoided, and we can convert
(\ref{eq.467654765b}) in a form analogue to (\ref{eq.decoS}). To see
this heuristically, remark that, for large $n$,
\be
\begin{split}
\sum_{y=1}^n
\ln \Big( 1-\frac{z y}{n} \Big)
&\simeq
n
\int_0^1 \rd y
\ln ( 1-z y )
=
n
\left( -1-\frac{(1-z)\ln(1-z)}{z} \right)
\ef.
\end{split}
\ee
We can transform the `$\simeq$' sign into a systematic hierarchy of
bounds, in inverse powers of $n$, using the customary Euler-Maclaurin
expansion (see e.g.\ \cite[Chapt.\ 7]{WWat}). 

Let us write this more explicitly for the case ``sum of logs'' that
often occurs in saddle point integrals associated to specifiable
combinatorial structures.  Let $f(z,x)$ a smooth function of two
variables, let $f'$, $f''$,\ldots denote differentiation w.r.t.\ the
second argument, and let $g(z,x)=f'(z,x)/f(z,x)$. We have
\be
\begin{split}
\sum_{y=1}^n
\ln f(z,y/n)
& \in
n
\int_0^1 \rd y
\ln f(z,y)
+
\ln \sqrt{\frac{f(z,1)}{f(z,0)}}
\\ & \quad
+
\sum_{k=1}^{k_{\rm Max}-1}
n^{-2k+1}
\frac{(-1)^{k-1} B_k}{(2k)!}
\left(g^{(2k-2)}(z,1) - g^{(2k-2)}(z,0) \right)
\\ & \quad
\ptil
n^{-2k_{\rm Max}+2}
\frac{2 \zeta(2n)}{(2 \pi)^{2n}}
\int_0^1
\rd y
\,
|g^{(2 k_{\rm Max}-1)}(z,y)|
\ef;
\end{split}
\ee
(where the $B_j$'s are the \emph{Bernoulli numbers}). If one has even
a moderate control on $|g^{(k)}(z,y)|$ for $y \in [0,1]$, an estimate
in this form is easily integrated in the general construction of
the bounds performed in Section~\ref{sec.HieraBou}.

\section{A criterium for tails pruning}
\label{sec.tails}

Here we present a simple criterium for bounding the contour integral on
the open path $\gamma_{\rm rest}$, discussed in subection
\ref{sec.HieraBou}. We also discuss a very elementary bound on the
``tail completion'', the extra term arising from the fact that we
replace the Gaussian integrals on a finite interval by the complete
integral, by adding and subtracting a correction term.

Let us first consider the integral over $\gamma_{\rm rest}$.
Suppose that $\overline{S(z)}=S(\overline{z})$, and $A(z)=z^j$. 
In this case $x_{\pm}$ are complex conjugates.
Suppose that, at the radius $\rho=|z_*+x_+|$, $\reof \; S(\rho \exp(i
\theta))$ is monotone for $\theta \in [\mathrm{arg}(z_*+x_+),\pi]$. In
such a case we have
\be
\left|
\int_{\gamma_{\rm rest}}
\frac{\rd z}{2 \pi i z}
A(z) \,
\exp \big( n (S(z)-S(z_*) \big)
\right|
\leq
\rho^{j} \exp \big( n \; \reof (S(z_*+x_+)-S(z_*)) \big)
\ef.
\ee
For the tails of the Gaussian integrals, any customary bound on the
\emph{erf} error function makes the game. A good compromise between
simplicity and tightness is based on the following calculation, valid
for all $k \in \bbN$ and $a \in \bbR^+$
\be
\begin{split}
\int_{a}^{\infty} 
\!\!\!
\rd x \,x^k e^{-\frac{x^2}{2}}
&
=
\int_{0}^{\infty} 
\!\!\!
\rd x \,(a+x)^k e^{-\frac{a^2}{2}-ax-\frac{x^2}{2}}
\leq
e^{-\frac{a^2}{2}}
\int_{0}^{\infty} 
\!\!\!
\rd x \,(a+x)^k e^{-ax}
\\
&=
e^{-\frac{a^2}{2}}
\sum_{h=0}^k \binom{k}{h}
a^{k-h}
\cdot 
h!\, a^{-h-1}
\leq
a^{k-1}
e^{-\frac{a^2}{2}}
\sum_{h=0}^k 
\left( \frac{k}{a^2} \right)^h
\leq
a^{k-1}
e^{-\frac{a^2}{2}}
R
\ef,
\end{split}
\ee
where $R$ may be chosen to be $\max \big( k, (k/a)^2 \big)$, or also,
if $k<a^2$, extending the geometric sum to infinity, $(1-k/a^2)^{-1}$.
For $a \gg 1$ and $k=\cO(1)$, as in our application (where $a \sim \sqrt{N}$), 
the second estimate is the more tight.

\end{document}